\documentclass[10pt]{article}
\usepackage{amsfonts}
\usepackage{amsmath}
\usepackage{amssymb}
\usepackage{epsfig}
\usepackage{graphicx,color}% Include figure files
\usepackage{graphicx}
\def\nslash{n\!\!\!\slash}
\def\vslash{v\!\!\!\slash}
\def\pslash{p\!\!\!\slash}
\def\ks{K^*_0}
\def\kp{K^{*+}_0}
\begin{document}

\title{Study of Decay Modes $B \to K_0^*(1430)\phi$}
\author{~~C.~S Kim,$^1$~\footnote{Email: ~~cskim@yonsei.ac.kr}~~~~  Ying Li,$^{1,2,4}$~\footnote{Email: ~~liying@ytu.edu.cn}~~~~ Wei
Wang$^3$~\footnote{Email: ~~wei.wang@ba.infn.it}
 \\\small{\it  1.Department of Physics, Yonsei University, Seoul
120-479, Korea }\\
\small {\it  2. Department of Physics, Yantai University, Yantai
264-005,
China }\\
\small {\it  3.Istituto Nazionale di Fisica Nucleare, Sezione di Bari, Bari 70126, Italy }\\
\small {\it  4. Kavli Institute for Theoretical Physics China
(KITPC), Beijing,100-080, China} } \maketitle
\begin{abstract}

\end{abstract}
Within the framework of perturbative QCD approach based on
$\mathbf{k_T}$ factorization, we investigate the charmless decay
mode $B \to K_0^*(1430)\phi$. Under two different scenarios (S1 and
S2) for the description of scalar meson $K_0^*(1430)$, we explore the branching
fractions and related $CP$ asymmetries. Besides the dominant
contributions from the factorizable emission diagrams, penguin
operators in the annihilation diagrams could also provide considerable
contributions. The central values of our predictions are larger than
those from the QCD factorization in both scenarios. Compared with the
experimental measurements of the BaBar collaboration, the result
of neutral channel in the S1 agrees with experimental data, while
the result of the charged one is a bit smaller than the data. In the
S2 scenario, although the central value for the branching fractions of both
channels are much larger than the data, the predictions could agree
with the data due to the large uncertainties to the branching
fractions from the hadronic input parameters. The $CP$ asymmetry in
the charged channel is small and not sensitive to CKM angle
$\gamma$. With the accurate data in near future from the various $B$ factories, these
predictions will be under stringent tests.
~
\newpage
%-------------------------------------------------------------------------------
%                               Introduction
%===============================================================================
\section{Introduction}
%===============================================================================
The $b \to ss\bar s$ transition, inducing many  non-leptonic
charmless $B$ meson decay processes such as $B \to  K_S\phi$, $B \to
K_S\eta(\eta^\prime) $ and $B \to K^*\phi $, has attracted much
interest because it serves as an ideal platform to probe the
possible new physics (NP) beyond the standard model (SM). However,
the kind of transition involving a scalar meson have more
ambiguities due to intriguing but mysterious underlying nature of
scalar mesons. In the spectroscopy study, there are two different
scenarios to describe the scalar mesons. The scenario-1 (S1) is the
naive 2-quark model: the nonet mesons below 1 GeV are treated as the
lowest lying states, and the ones near 1.5 GeV are the first
orbitally excited state. In the scenario-2 (S2), the nonet mesons
near 1.5 GeV are viewed as the lowest lying states, while the mesons
below 1 GeV may be viewed as exotic states beyond the quark model
such as four-quark bound states. Under these two pictures, many
$B\to SP$ modes, such as $B \to f_0K$, induced by $b\to s s\bar s$
transition have been calculated in both QCD factorization (QCDF)
approach~\cite{Cheng:2005ye,Cheng:2005nb} and perturbative QCD
(PQCD) approach
\cite{Chen:2002si,Wang:2006ria,Shen:2006ms,Liu:2009xm}. Within
proper regions for the input parameters, many theoretical results
could agree with the experimental data.

In this work, we will study the $B\to  K^*_0(1430)\phi$ decays in
the perturbative QCD approach~\cite{Lu:2000em}.  On the experimental
side, the branching ratios of $B \to K_0^*(1430)\phi$ have been
measured with good precision \cite{Aubert:2006uk,:2008zzd}:
\begin{eqnarray}
\label{eq:data-1}
 {\cal B}(\overline B^0\to \overline K_0^{*0}(1430)\phi) &=& (4.6\pm0.7\pm0.6)\times
 10^{-6}~, \\
 {\cal B}( B^\pm\to  K_0^{*\pm}(1430)\phi) &=& (7.0\pm1.3\pm0.9)\times
 10^{-6}~,
\end{eqnarray}
where the result for the neutral channel has been updated
as~\cite{Aubert:2008zz}
\begin{eqnarray}
 {\cal B}(\overline B^0\to \overline K_0^{*0}(1430)\phi) &=& (3.9\pm0.5\pm0.6)\times
 10^{-6}~.
\end{eqnarray}
Compared with the $B\to K\phi$ decay~\cite{Barberio:2006bi}
\begin{eqnarray}
 {\cal B}(\overline B^0\to \overline K^0 \phi) &=& (8.3^{+1.2}_{-1.0})\times
 10^{-6}~, \\
 {\cal B}( B^\pm\to  K^{\pm}\phi) &=& (8.30\pm0.65)\times
 10^{-6}~,
\end{eqnarray}
we can see that the decay channels with a
scalar meson in the final state, $B \to K_0^*(1430)\phi$, seem to have a bit smaller branching
fractions. In Refs.~\cite{Chen:2005cx,Chen:2007qj}, the decay
$\overline B^0\to \overline K_0^{*0}(1430)\phi$ has been studied
within the framework of generalized factorization in which the
non-factorizable effects are described by the parameter $N_c^{\rm
eff}$, the effective number of colors. The predicted branching ratio (BR) varies from
$10^{-7}$ to $10^{-5}$, depending on the different values for
$N_c^{\rm eff}$. Without the information for non-factorizable
effects, one cannot make a precise prediction of the BR.
The QCDF calculation of this and other
modes has also been presented in Ref. \cite{Cheng:2007st}, and the
predicted central value of ${\cal B}(\overline B^0\to\overline
K^{*0}_0(1430)\phi)$ deviates from the experimental data, though it
can be accommodated within large theoretical errors. It is necessary
to analyze these channels in the PQCD approach with different
treatments for the matrix elements of the four-quark operators,
which is helpful to probe the structure of the scalar meson
model-independently.

%Annihilation diagrams play an important role in the explanation of
%the branching ratios, direct CP asymmetries in $B\to\pi\pi,\pi K$
%decays and polarizations of $B\to VV$ decays. Through keeping the
%transverse momentum $k_T$, one can smear the endpoint singularity
%effectively in the PQCD approach. The annihilation diagram could be
%directly evaluated and one can make direct predictions to the
%branching fractions.

The layout of the present paper is as follows: In Sec. \ref{sec:2} we
introduce the input parameters including the decay constants and
light-cone distribution amplitudes. The factorization formulae in
the perturbative QCD approach are given in Sec. 3. Numerical results
and discussions are presented in Sec. 4. Summary of this work is also
given in Sec. 4.
%-------------------------------------------------------------------------------
%                              Input Parameters
%===============================================================================
\section{Input Parameters}\label{sec:2}
%===============================================================================

In the $B$ meson rest frame, the $B$ meson momentum $P_1$, the
$\phi$ meson momentum $P_2$, the longitudinal polarization vector
$\epsilon_L$, and the kaon momentum $P_3$ are chosen, in light-cone
coordinates, as
\begin{eqnarray}
P_1=\frac{M_B}{\sqrt{2}}(1,1,{\bf 0}_T)\;,\;\;\;
P_2&=&\frac{M_B}{\sqrt{2}}(1-r_{K_0^*}^2,r_\phi^2,{\bf
0}_T)\;,\;\;\; P_3=\frac{M_B}{\sqrt{2}}(r_{K_0^*}^2,1-r_\phi^2,{\bf
0}_T)\;,\;\;\;
\nonumber \\
\epsilon &=&\frac{1}{\sqrt{2}r_\phi}(1-r_{K_0^*}^2,-r_\phi^2,{\bf
0}_T)\;, \label{pa}
\end{eqnarray}
with the ratio $r_{\phi(K_0^*)}=m_{\phi({K_0^*})}/M_B$, and $m_\phi$,
$m_{K_0^*}$ being the $\phi$ meson mass  and $K_0^*$ meson mass,
respectively. The momentum of the light antiquark in the $B$ meson
and the light quarks in the final mesons are denoted as $k_1$, $k_2$
and $k_3$ respectively. Using the intrinsic variables (momentum
fractions and the transverse momentum), we can choose
\begin{eqnarray}
k_1 = (0,x_1 P_1^-,{\bf k}_{1T}), \quad k_2 = (x_2 P_2^+,0,{\bf
k}_{2T}), \quad k_3 = (0,x_3 P_3^-,{\bf k}_{3T}).
\end{eqnarray}

The decay constants of scalar meson are defined by
\begin{eqnarray}
 \langle S(p)|\bar
 q_2\gamma_{\mu}q_1|0\rangle=f_Sp_{\mu}\;,\; \langle S|\bar q_2q_1|0\rangle=m_S\bar f_S,
\end{eqnarray}
where the decay constant $f_S$ of the vector current and $\bar f_S$ of
the scalar current are related by equations of motion $\mu_sf_S=\bar
f_S$, with $\mu_s=\frac{m_S}{m_2(\mu)-m_1(\mu)}$. The parameter $m_S$ is the mass
of the scalar meson, and $m_1$, $m_2$ are the running current quark
masses. Inputs of the scalar mesons in our calculation, including
the decay constants, running quark masses and the Gegenbauer moments
defined in the following, are quoted from Ref.~\cite{Cheng:2005nb}.

For the scalar meson wave function, the twist-2 light-cone
distribution amplitude (LCDA) $\phi_S(x)$ and twist-3 LCDAs
$\phi_S^s(x)$ and $\phi_S^{\sigma}$ for the scalar mesons can be
combined into a single matrix element:
\begin{eqnarray}
 \langle K_0^{*+}(p)|\bar{u}_{\beta}(z)s_{\alpha}(0)|0\rangle
&=&\frac{1}{\sqrt{6}}\int^1_0dxe^{ixp \cdot z}\bigg\{ \pslash
\phi_{K^{*+}_0}(x)+ m_S\phi^S_{K^{*+}_0}(x)+\frac{1}{6}
m_S\sigma_{\mu\nu}p^{\mu}z^{\nu}\phi^{\sigma}_{K^{*+}_0}(x)\bigg\}_{\alpha\beta}\nonumber\\
&=& \frac{1}{\sqrt{6}}\int^1_0dxe^{ixp \cdot z}\bigg\{ \pslash
\phi_{K^{*+}_0}(x)+ m_S\phi^S_{K^{*+}_0}(x)+
m_S(\nslash\vslash-1)\phi^T_{K^{*+}_0}(x)\bigg\}_{\alpha\beta},
\end{eqnarray}
where $v$ and $n$ are dimensionless vectors on the light cone, and
$n$ is parallel with the moving direction of the scalar meson.
 The distribution amplitudes $\phi_{K^*_0}(x)$,
$\phi^S_{K^*_0}(x)$ and $\phi^{\sigma}_{K^*_0}(x)$ are normalized
as:
\begin{eqnarray} \int^1_0dx\phi_{K^*_0}(x)=\frac{f_{K^*_0}}{2\sqrt{6}},\,\,\,\,\,\,
\int^1_0dx\phi^S_{K^*_0}(x)=\int^1_0dx\phi^{\sigma}_{K^*_0}(x)=\frac{\bar{f}_{K^*_0}}{2\sqrt{6}},
\end{eqnarray} and
$\phi^T_{K^*_0}(x)=\frac{1}{6}\frac{d}{dx}\phi^{\sigma}_{K^*_0}(x)$.
For the $K^{*+}_0$ meson, the decay constant $f_{K^*_0}$ has the
opposite sign with that of the $K^{*-}_0$ meson.

Under the conformal spin symmetry, the twist-2 LCDA
$\phi_{K^*_0}(x)$ can be expanded as:
\begin{eqnarray}
 \phi_{K^*_0}(x,\mu)\nonumber
&=&\frac{\bar{f}_{K^*_0}(\mu)}{2\sqrt{6}}6x(1-x)\bigg[B_0(\mu)
+\sum\limits^{\infty}_{m=1}B_m(\mu)C_m^{3/2}(2x-1)\bigg]\\
&=&-\frac{{f}_{K^*_0}(\mu)}{2\sqrt{6}}6x(1-x)\bigg[-1+\mu_S\sum\limits^{\infty}_{m=1}
B_m(\mu)C_m^{3/2}(2x-1)\bigg],
\end{eqnarray}
where $B_m(\mu)$ and $C_m^{3/2} (x)$ are the Gegenbauer moments and
Gegenbauer polynomials, respectively. The Gegenbauer moments $B_1$,
$B_3$ of distribution amplitudes for $K^*_0 $ and the decay
constants have been calculated in the QCD sum
rules~\cite{Cheng:2005nb} as
\begin{eqnarray}
\mbox{S} \,{\rm 1}&:&\,\,\,\, B_1=0.58\pm 0.07,\;\;\;\;\;\;
B_3=-1.20\pm 0.08,\;\;\;\;\;\; \bar{f}_{K^*_0}(1 \mathrm{GeV})=-(300 \pm 30)~~\mathrm{MeV};\nonumber\\
\mbox{S} \,{\rm 2}&:&\,\,\,\, B_1=-0.57\pm 0.13,\;\;\;\;
B_3=-0.42\pm 0.22,\;\;\;\;\;\; \bar{f}_{K^*_0}(1 \mathrm{GeV})=(445
\pm 50)~~ \mathrm{MeV}.
\end{eqnarray}
All the above values are  taken at $\mu=1$ GeV.

For the twist-3 LCDAs, they have been promoted in the
Ref.~\cite{Lu:2006fr} with large uncertainties, so we take the
asymptotic form in our numerical calculation for simplicity:
\begin{eqnarray}
\phi^S_S(x)=\frac{\bar{f}_S}{2\sqrt{6}},\,\,\,\,\,\phi^T_S(x)=\frac{\bar{f}_S}{2\sqrt{6}}(1-2x).
\end{eqnarray}
Up to twist-3 accuracy, the vector meson's wave functions are
collected as
\begin{eqnarray}
\langle \phi(P_2,\epsilon_L)|\bar s_{\beta}(z) s_{\alpha}
(0)|0\rangle &=&\frac{1}{\sqrt{6}}\int_0^1 dx e^{ixP_2\cdot z}
\left[m_\phi\not\! \epsilon^*_L \phi_\phi(x) +\not\!
\epsilon^*_L\not\! P_2 \phi_{\phi}^{t}(x) +m_\phi
\phi_\phi^s(x)\right]_{\alpha\beta}, \label{lpf}
\end{eqnarray}
for longitudinal polarization.  The distribution amplitudes can be
parametrized as:
\begin{eqnarray}
\phi_{\phi}(x)&=&\frac{3f_{\phi}}{\sqrt{6}} x (1-x)\left[1+
a_{2\phi}^{||}C_2^{3/2} (2x-1)\right],\;\nonumber\\
\phi^t_\phi(x) &=& \frac{3f^T_\phi}{2\sqrt 6}(2x-1)^2,\nonumber\\
 \phi^s_\phi(x)&=&\frac{3f_\phi^T}{2\sqrt 6} (1-2x)~,\label{phiphi}
\end{eqnarray}
with the Gegenbauer coefficient $a_{2\phi}^{||}(1{\rm
GeV})=0.18\pm0.08$ \cite{Ball:2007rt}.

Since the $B$ meson is a pseudo-scalar heavy meson, the structure
$(\gamma^\mu\gamma_5)$ and $\gamma_{5}$ components remain as leading
contributions. Then, $\Phi_{B}$ is written by
\begin{equation}
 \Phi_{B} = \frac{i}{\sqrt{6}}
\left\{ (\not \! P_B \gamma_5) \phi_B^A + \gamma_{5} \phi_B^P
\right\},
\end{equation}
where $P_B$ is the corresponding meson's momentum, and
$\phi_B^{A,P}$ are Lorentz scalar distribution amplitudes. As heavy
quark effective theory leads to $\phi_B^P \simeq M_B \phi_B^A$, $B$
meson's wave function can be expressed by
\begin{equation}
 \phi_{B}(x,b) = \frac{i}{\sqrt{6}}
\left[ (\not \! P_B \gamma_5) + M_B \gamma_5 \right] \phi_B(x,b).
\end{equation}
For the $B$ meson distribution amplitude, we adopt the model:
\begin{eqnarray}
\phi_{B}(x,b)=N_{B}x^{2}(1-x)^{2}\exp \left[ -\frac{1}{2} \left(
\frac{xM_{B}}{\omega _{B}}\right) ^{2} -\frac{\omega
_{B}^{2}b^{2}}{2}\right] \label{bw} \;,
\end{eqnarray}
with the shape parameter $\omega_{B}=0.4$ GeV, which has been tested
in many channels such as $B\to \pi\pi, K\pi$ \cite{Lu:2000em}. The
normalization constant $N_{B}= 91.784$ GeV is related to the decay
constant $f_{B}=190$ MeV. In the above model, $\phi_B$ has a sharp
peak at $x\sim \bar\Lambda/M_B\sim 0.1$.

%-------------------------------------------------------------------------------
%                              Input Parameters
%===============================================================================
\section{Analytical Formulae}
%===============================================================================

In the PQCD approach, after the integration over $k_1^+$, $k_2^+$,
and $k_3^-$, the decay amplitude for $B \to {K^*_0}  \phi$ decay can
be conceptually written as
\begin{eqnarray} {\cal A} &\sim &\int\!\! d x_1 d x_2 d
x_3 b_1 d b_1 b_2 d b_2 b_3 d b_3 \nonumber\\
 && ~~\times \mathrm{Tr}
\left [ C(t) \Phi_{B}(x_1,b_1) \Phi_{\phi}(x_2,b_2)
\Phi_{K^*_0}(x_3, b_3) H(x_i, b_i, t) S_t(x_i)\, e^{-S(t)} \right ],
\label{eq:a2}
\end{eqnarray}
where $x_i$ are momenta fraction of light quarks  in each
meson. $\mathrm{Tr}$ denotes the trace over Dirac and color indices,
$C(t)$ is the Wilson coefficient evaluated at scale $t$, and the hard
kernel $H(k_1,k_2,k_3,t)$ is the hard part and can be calculated
perturbatively. And the function $\Phi_M$ is the wave function, the
function $S_t(x_i)$ describes the threshold resummation which smears
the end-point singularities on $x_i$, and the last term,
$e^{-S(t)}$, is the Sudakov form factor which suppresses the soft
dynamics effectively.

%%%%%%%%%%%%%%%%%%%%%%%%%%%%%%%%%%%%%%%%%%%%%%%%%%%%%%%%%%%%%%%%%%%%%%
\begin{figure}[tb]
\begin{center}
\psfig{file=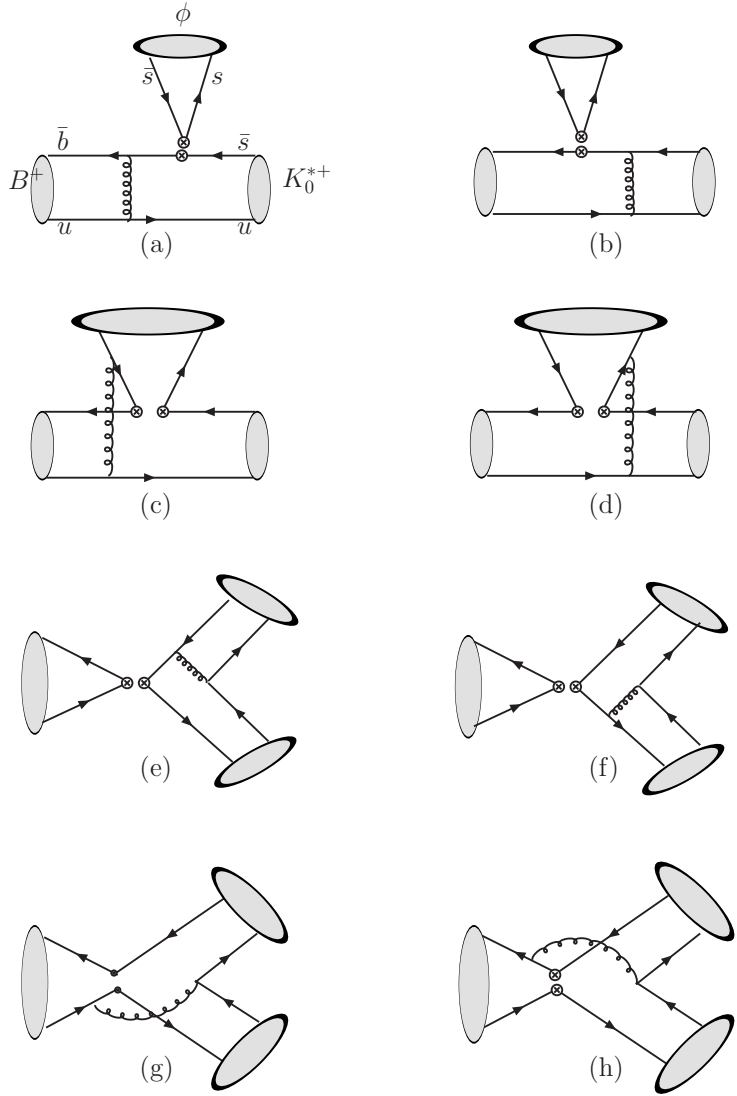,width=9.7cm,angle=0}
\end{center}
\caption{The leading order Feynman diagrams for $B^+\to \kp \phi$
decay in PQCD approach }\label{Feynman}
\end{figure}
%%%%%%%%%%%%%%%%%%%%%%%%%%%%%%%%%%%%%%%%%%%%%%%%%%%%%%%%%%%%%%%%%%%%%%%

In the standard model, the effective weak Hamiltonian mediating
flavor-changing neutral current transitions of the type $b\to s $
has the form:
\begin{eqnarray}\label{hamiton}
 {\cal H}_{eff}={G_F\over \sqrt 2}\Big[\sum\limits_{p=u,c}V_{pb}V^*_{ps}\Big(
 C_1O_1^p+C_2O_2^p\Big)-V_{tb}V^*_{ts}\sum\limits_{i=3}^{10,7\gamma,8g}
 C_iO_i\Big],
\end{eqnarray}
where the explicit form of the operator $O_i$ and the corresponding
Wilson coefficient $C_i$ can be found in Ref. \cite{Buras}.
$V_{p(t)b}$, $V_{p(t)s}$ are the  CKM  matrix elements. According to
effective Hamiltonian (\ref{hamiton}), we draw the lowest order
diagrams of this channel in Fig. \ref{Feynman}.

We first calculate the usual factorizable emission diagrams (a) and
(b). If we insert the $(V-A)(V-A)$ or $(V-A)(V+A)$ operators in the
corresponding vertexes, the amplitude associated to these currents
is given as:
\begin{multline}
F_e =-8\pi C_F m_B^4f_\phi \int_0^1 dx_1 dx_3 \int_0^{\infty}
b_1db_1\, b_2db_3\, \phi_B(x_1,b_1)\\\bigg\{ \left[
(1+x_3)\phi_{\ks}(x_3)+r_{\ks}(1-2x_3) \left(
\phi_{\ks}^S(x_3)+\phi_{\ks}^T(x_3) \right) \right] a(t_a)
E_{e}(t_a)h_{e}(x_1,x_3,b_1,b_3)
\\
 +2r_{\ks} \phi_{\ks}^S({x_3})a(t_b)E_{e}(t_b)
h_{e}(x_3,x_1,b_3,b_1) \bigg\}.\label{fe}
\end{multline}
In the above formulae, $C_F=4/3$ is the group factor of the
$SU(3)_{c}$ gauge group. We will use the same conventions for the
functions $h_e$ and $E_{e}(t')$ including the Sudakov factor and jet
function as those in Ref.~\cite{Ali:2007ff}. The $(S-P)(S+P)$
operator does not contribute to this decay as the emission particle
is a vector particle. For the non-factorizable diagrams (c) and (d),
all three meson wave functions are involved. For the $(V-A)(V-A)$
operators, the result can be read as:
 \begin{multline}
M_{e}^{LL} = \frac{32 \pi}{ \sqrt{2N_C}}C_Fm_B^4 \int_0^1
dx_1dx_2dx_3 \int_0^{\infty} b_1 db_1\, b_2 db_2\,\phi_B(x_1,b_1)
\phi_{\phi}(x_2)\\
\bigg\{
\left[(x_2-1)\phi_{\ks}(x_3)+r_{\ks}x_3\bigg(\phi_{\ks}^S(x_3)-\phi_{\ks}^T(x_3)\bigg)\right]
 a(t_c)E'_{e}(t_c)h_n(x_1,1-x_2,x_3,b_1,b_2)
 \\
+\left[(x_3+x_2)\phi_{\ks}(x_3) -r_{\ks} x_3
\left(\phi_{\ks}^S(x_3)+\phi_{\ks}^T(x_3)\right)\right] a(t_d)
E'_{e}(t_d) h_n(x_1,x_2,x_3,b_1,b_2) \bigg\}\;.\label{me1}
 \end{multline}
For $(V-A)(V+A)$ and the $(S-P)(S+P)$ operators, the formulae are
listed as:
\begin{multline}
M_{e}^{LR} = \frac{32 \pi}{ \sqrt{2N_C}}C_Fm_B^4\int_0^1 dx_1dx_2
dx_3\int_0^{\infty} b_1 db_1\, b_2 db_2\,\phi_B(x_1,b_1)
r_\phi\\\bigg\{
\bigg[(1-x_2)\phi_{\ks}(x_3)(\phi_\phi^s(x_2)+\phi_\phi^t(x_2))+r_{\ks}
\bigg(\phi_\phi^s(x_2)\left[(x_3-x_2+1)\phi_{\ks}^S(x_3)+(x_3+x_2-1)\phi_{\ks}^T(x_3)\right]\\
-\phi_\phi^t(x_2)\left[(x_3+x_2-1)\phi_{\ks}^S(x_3)+(x_3-x_2+1)\phi_{\ks}^T(x_3)\right]\bigg)\bigg]
 a(t_c)E'_{e}(t_c)h_n(x_1,1-x_2,x_3,b_1,b_2)
 \\
+\bigg[x_2\phi_{\ks}(x_3)(\phi_\phi^t(x_2)-\phi_\phi^s(x_2))-r_{\ks}
\bigg(\phi_\phi^s(x_2)\left[(x_3+x_2)\phi_{\ks}^S(x_3)+(x_3-x_2)\phi_{\ks}^T(x_3)\right]\\
+\phi_\phi^t(x_2)\left[(x_3-x_2)\phi_{\ks}^S(x_3)+(x_3+x_2)\phi_{\ks}^T(x_3)\right]\bigg)\bigg]
a(t_d) E'_{e}(t_d) h_n(x_1,x_2,x_3,b_1,b_2) \bigg\}\;,\label{me2}
 \end{multline}
\begin{multline}
 {\cal M}_{e}^{SP} = -\frac{32 \pi}{ \sqrt{2N_C}}C_Fm_B^4 \int_0^1
dx_1dx_2dx_3 \int_0^{\infty} b_1 db_1\, b_2 db_2\,\phi_B(x_1,b_1)
\phi_{\phi}(x_2)\\\bigg\{
\left[(1-x_2+x_3)\phi_{\ks}(x_3)-r_{\ks}x_3\bigg(\phi_{\ks}^S(x_3)+\phi_{\ks}^T(x_3)\bigg)\right]
 a(t_c)E'_{e}(t_c)h_n(x_1,1-x_2,x_3,b_1,b_2)
 \\
+\left[-x_2\phi_{\ks}(x_3) +r_{\ks} x_3
\left(\phi_{\ks}^S(x_3)-\phi_{\ks}^T(x_3)\right)\right] a(t_d)
E'_{e}(t_d) h_n(x_1,x_2,x_3,b_1,b_2) \bigg\}\;,\label{me3}
 \end{multline}

Diagrams (e) and (f)  are the factorizable annihilation diagrams, and
the $(V-A)(V-A)$ kind of operators' contributions are
\begin{multline}
F^L_{a}(a) = -8\pi C_F m_B^4f_B \int_0^1  dx_2dx_3 \int_0^{\infty}
b_2db_2\, b_3db_3\
 \\
 \times \Bigg\{\Big[(x_3-1)\phi_{\ks}(x_3)\phi_{\phi}(x_2) -
2r_{\phi} r_{\ks}\left((x_3-2) \phi_{\ks}^S(x_3)
 -x_3\phi_{\ks}^T(x_3)\right)\phi_\phi^s(x_2)\Big] a(t_e)
E_{a}(t_e) h_{a}(x_2,1-x_3, b_2, b_3)
\\
 + \Big[x_2\phi_{\ks}(x_3)\phi_{\phi}(x_2) -
2r_{\phi} r_{\ks}\phi_{\ks}^S(x_3)\left((x_2+1) \phi_{\phi}^s(x_2)
 +(x_2-1)\phi_{\phi}^t(x_2)\right)\Big]
 a(t_f) E_{a}(t_f) h_{a}(1-x_3,x_2, b_3,
b_2)\Bigg\},\label{fa1}
\end{multline}
and the result from $(S-P)(S+P)$ currents is:
\begin{multline}
 F^{SP}_{a}(a) =16\pi C_F m_B^4f_B\int_0^1  dx_2dx_3 \int_0^{\infty}
b_2db_2\, b_3db_3\
 \\
 \times \Bigg\{\Big[2r_\phi\phi_{\ks}(x_3)\phi_{\phi}^s(x_2) +
r_{\ks}(x_3-1)\left(\phi_{\ks}^S(x_3)
 +\phi_{\ks}^T(x_3)\right)\phi_\phi(x_2)\Big] a(t_e)
E_{a}(t_e) h_{a}(x_2,1-x_3, b_2, b_3)
\\
-\Big[2r_{\ks}\phi_{\ks}^S(x_3)\phi_{\phi}(x_2) +
r_{\phi}x_2\left(\phi_{\phi}^t(x_2)
 -\phi_{\phi}^s(x_2)\right)\phi_{\ks}(x_3)\Big]
 a(t_f) E_{a}(t_f) h_{a}(1-x_3,x_2, b_3, b_2)\Bigg\}.\label{fa2}
\end{multline}

The last two diagrams in Fig. \ref{Feynman} are the non-factorizable
annihilation diagrams, whose contributions are
\begin{multline}
 M^{LL}_{a}(a) =\frac{32 \pi}{ \sqrt{2N_C}}C_Fm_B^4 \int_0^1
dx_1dx_2dx_3 \int_0^{\infty} b_1 db_1\, b_2 db_2\,\phi_B(x_1,b_1)\\
 \bigg\{ \bigg[ x_2 \phi_{\ks}(x_3)\phi_{\phi}(x_2)
 +r_{\phi}r_{\ks} \phi_{\phi}^s(x_2) \left(
(x_3-x_2-3)\phi_{\ks}^S(x_3)+(x_3+x_2-1) \phi_{\ks}^T(x_3) \right)
\\ - r_{\phi}r_{\ks} \phi_{\phi}^t(x_2) \left(
(x_3+x_2-1)\phi_{\ks}^S(x_3)+(x_3-x_2+1) \phi_{\ks}^T(x_3)
\right)\bigg]a(t_g)E'_{e}(t_g) h_{na}(x_1,x_3,x_2,b_1,b_2)\\ +
\bigg[(x_3-1) \phi_{\ks}(x_3)\phi_{\phi}(x_2)
 -r_{\phi}r_{\ks} \phi_{\phi}^t(x_2) \left(
(x_3+x_2-1)\phi_{\ks}^S(x_3)+(-x_3+x_2+1) \phi_{\ks}^T(x_3) \right)
\\ - r_{\phi}r_{\ks} \phi_{\phi}^s(x_2) \left(
(x_3-x_2-1)\phi_{\ks}^S(x_3)-(x_3+x_2-1) \phi_{\ks}^T(x_3)
\right)\bigg] a(t_h)E'_{e}(t_h) h'_{na}(x_1,x_3,x_2,b_1,b_2)
\bigg\}, \label{ma1}
\end{multline}
\begin{multline}
 M^{LR}_{a}(a) =\frac{32 \pi}{ \sqrt{2N_C}}C_Fm_B^4 \int_0^1
dx_1dx_2dx_3 \int_0^{\infty} b_1 db_1\, b_2 db_2\,\phi_B(x_1,b_1)\\
 \bigg\{ \bigg[ (2-x_2) r_\phi \phi_{\ks}(x_3)(\phi_{\phi}^s(x_2)+\phi_{\phi}^t(x_2))
 +(x_3+1) r_{\ks} (\phi_{\ks}^S(x_3)-\phi_{\ks}^T(x_3))\phi_{\phi}(x_2)
 \bigg]a(t_g)E'_{e}(t_g) h_{na}(x_1,x_3,x_2,b_1,b_2)\\
 + \bigg[ x_2 r_\phi \phi_{\ks}(x_3)\left(\phi_{\phi}^s(x_2)+\phi_{\phi}^t(x_2)\right)
 -(x_3-1) r_{\ks} \left(\phi_{\ks}^S(x_3)-\phi_{\ks}^T(x_3)\right)\phi_{\phi}(x_2)\bigg]
 a(t_h)E'_{e}(t_h) h'_{na}(x_1,x_3,x_2,b_1,b_2)
\bigg\}.\label{ma2}
\end{multline}

By combining the contributions from different diagrams with
corresponding Wilson coefficients, one obtains the total decay
amplitudes as
\begin{eqnarray}
{\cal A}(\overline B\to \overline K_0^{*0}(1430)\phi) &=&
V_{tb}^{*}V_{ts} \Bigg\{ F_{e}\left[a_3+a_4+a_5-\frac{1}{2}
(a_7+a_9+a_{10})\right]
  \nonumber
 \\
 &&
 + M_{e}^{LL} \left[C_3+C_{4}-\frac{1}{2}C_9-\frac{1}{2}C_{10}\right]
 +M_{e}^{LR} \left[C_5-\frac{1}{2}C_{7}\right]+ M_{e}^{SP}
 \left[C_6-\frac{1}{2}C_{8}\right]
    \nonumber
 \\
 &&
   +F_{a}^{LL}\left[a_{4}-\frac{1}{2}a_{10}\right]
   + F_{a}^{SP}\left[a_{6}-\frac{1}{2}a_{8}\right]\nonumber
 \\
&& + M_{a}^{LL}\left[C_3-\frac{1}{2}C_9 \right] +
M_{a}^{LR}\left[C_5-\frac{1}{2}C_{7}\right]\Bigg\};
\end{eqnarray}
\begin{eqnarray}
{\cal A}(B^+\to  K_0^{+*}(1430)\phi) &=& V_{tb}^{*}V_{ts} \Bigg\{
F_{e}\left[a_3+a_4+a_5-\frac{1}{2} (a_7+a_9+a_{10})\right]
  \nonumber
 \\
 &&
 + M_{e}^{LL} \left[C_3+C_{4}-\frac{1}{2}C_9-\frac{1}{2}C_{10}\right]
 +M_{e}^{LR} \left[C_5-\frac{1}{2}C_{7}\right]+ M_{e}^{SP}
 \left[C_6-\frac{1}{2}C_{8}\right]
    \nonumber
 \\
 &&
+F_{a}^{LL}\left[a_{4}+a_{10}\right]+
F_{a}^{SP}\left[a_{6}+a_{8}\right]  +
M_{a}^{LR}\left[C_5+C_{7}\right]
 + M_{a}^{LL}\left[C_3+C_9 \right] \Bigg\}\nonumber\\
 &&
 -V_{ub}^{*}V_{us} \Bigg\{F_{a}^{LL}\left[C_2+\frac{1}{3}C_{1}\right]
 + M_{a}^{LL}C_1 \Bigg\},
\end{eqnarray}
where $C_i$ are the Wilson coefficients for the four-quark operators
and $a_i$ is defined as the combination of the Wilson coefficients:
\begin{eqnarray}
 a_i=C_i+\frac{C_{i\pm1}}{N_c}
\end{eqnarray}
for an odd (even) value of $i$. %The decay width is then expressed as
%\begin{equation}
% \Gamma = \frac{G_F^2 }{32\pi M_B}
%\left|\cal A\right|^2. \label{eq:width}
%\end{equation}
%-------------------------------------------------------------------------------
%                              Numerical Results
%--------------------------------------------------------------------------------
\section{Numerical Results}
%--------------------------------------------------------------------------------
The CKM phase $\gamma$ is defined via
\begin{eqnarray}
V_{ub}=|V_{ub}|e^{-i\gamma},
\end{eqnarray}
and the CKM matrix elements that we used in the calculation are
$|V_{ub}|=3.51\times 10^{-3}$, $|V_{us}|=0.225$,
$|V_{cb}|=41.17\times 10^{-3}$ and
$|V_{cs}|=0.973$~\cite{Charles:2004jd}. Moreover, we employ the unitary angle
$\gamma=70^{\circ}$, the masses $m_{B}=5.28$ GeV and $m_{\phi}=1.02$
GeV. The longitudinal decay constant of $\phi$ could be extracted
through the leptonic $\phi\to e^+e^-$ decay~\cite{PDG}
\begin{eqnarray}
\Gamma(\phi\to e^+e^-)=\frac{4\pi \alpha_{\rm em}^2e_s^2
f_{\phi}^2}{3m_{\phi}},
\end{eqnarray}
which gives
\begin{eqnarray}
 f_\phi&=&215~{\rm MeV}.
\end{eqnarray}
For the transverse decay constant, we use the recent Lattice QCD
result~\cite{Allton:2008pn} at 2~GeV
\begin{eqnarray}
\frac{f_\phi^T}{f_\phi}=0.750\pm0.008,
\end{eqnarray}
which corresponds to $f_{\phi}^{T}(1~\mathrm{GeV})=(178\pm2)$ MeV. The
${\bar{B}}_{d}^{0}$ ($B^{-}$) meson lifetime $\tau_{B^{0}}=1.530$ ps
($\tau_{B^{-}}=1.638$ ps) \cite{PDG}.

With the above input parameters, the $B\to K^{*}_0$ form factors are
given as
\begin{eqnarray}
& &F_1(q^2=0)=-0.42^{+0.04+0.03-0.09}_{-0.04-0.03+0.07},\,\,\,\,\,\,\,\,\,\,\,\,S1;\nonumber\\
&
&F_1(q^2=0)=~~0.73^{+0.08-0.10+0.15}_{{-0.08}+0.09-0.12},\,\,\,\,\,\,\,\,\,\,\,\,S2;
\end{eqnarray}
where the first two uncertainties are from decay constants and the distribution amplitudes
of the scalar meson, and the last uncertainty is from the $\omega_B$ in the
distribution amplitude of $B$ meson. The decay constant in S2 is
larger than that in S1, and contributions from the two terms
proportional to $B_1$ and $B_3$ are constructive in S2 but
destructive in S1. Thus the result for the form factor of $B\to
K^{*}_0$ in S2 is almost twice larger than that in S1.
 Compared with the previous
study of transition form factors~\cite{Li:2008tk}, we can see that
the present results for these form factors are a bit larger due to a
weaker suppression for the endpoint region from the jet function
$S_t(x)$.

The total decay amplitude for $B^+\to  K_0^{*+}(1430)\phi$ can be
written as:
\begin{equation}
{\cal A} = V_{ub}^*V_{us}T-V_{tb}^*V_{ts}P=V_{ub}^*V_{us}T[1+z
e^{i(\delta-\gamma)}],
\end{equation}
where $z=|V_{tb}^*V_{ts}/V_{ub}^*V_{us}||P/T|$ and $\delta$ is the
relative strong phase between tree diagrams ($T$) and penguin
diagrams ($P$). The decay width is expressed as:\begin{equation}
 \Gamma(B^+\to  K_0^{*+}(1430)\phi) = \frac{G_F^2 }{32\pi M_B}
|{\cal A}|^2=\frac{G_F^2 }{32\pi
M_B}|V_{ub}^*V_{us}T|^{2}[1+z^{2}+2z\cos(\delta-\gamma)].
\label{eq:width1}
\end{equation}
Similarly, we can get the decay width for $B^-\to
K_0^{*-}(1430)\phi$,
\begin{equation}
 \Gamma(B^-\to  K_0^{*-}(1430)\phi) = \frac{G_F^2 }{32\pi M_B}
|\overline{{\cal A}}|^2, \label{eq:width2}
\end{equation}
where
\begin{equation}
\overline {{\cal
A}}=V_{ub}V_{us}^{*}T-V_{tb}V_{ts}^{*}P=V_{ub}V_{us}^{*}T [1+z
e^{i(\delta+\gamma)}].
\end{equation}
{}From Eqs. (\ref{eq:width1}) and (\ref{eq:width2}), we get the averaged
decay width: \begin{eqnarray}
 \Gamma &=& \frac{G_F^2 }{32\pi M_B}
(|{\cal A}|^2/2+|\overline{\cal A}|^2/2)\hspace*{1cm}  \nonumber \\
&=&\frac{G_F^2 }{32\pi
M_B}|V_{ub}^*V_{us}T|^{2}[1+z^{2}+2z\cos\gamma\cos\delta].
\label{eq:width3}
\end{eqnarray}
Using Eqs. (\ref{eq:width1}) and (\ref{eq:width2}), the direct $CP$
violation  parameter is defined as
\begin{equation}
A_{CP}^{dir}=\frac{\Gamma(B^-\to K_0^{*-}(1430)\phi)-\Gamma(B^+\to
K_0^{*+}(1430)\phi)}{ \Gamma(B^-\to
K_0^{*-}(1430)\phi)+\Gamma(B^+\to K_0^{*+}(1430)\phi)}
=\frac{2z\sin\gamma\sin\delta}{1+2z\cos\gamma\cos\delta+z^{2}} .
\label{dcpv}
\end{equation}
Since only penguin operators work on the neutral decay mode, there
is no direct $CP$ asymmetry in the decay $B^0\to K_0^{*0}(1430)\phi$, and
its branching ratio can be calculated straightforwardly.

Using the parameters, we get the branching ratios in scenario 1
(S1):
 \begin{eqnarray}
  {\cal B}(B^0\to K_0^{*0}(1430)\phi)&=&3.7\times 10^{-6},\nonumber\\
{\cal B}( B^\pm\to K_0^{*\pm}(1430)\phi)&=&4.3\times 10^{-6},
\end{eqnarray}
while in scenario 2 (S2), the results are:
 \begin{eqnarray}
{\cal B}(B^0\to K_0^{*0}(1430)\phi)&=&23.6\times 10^{-6},\nonumber\\
{\cal B}( B^\pm\to K_0^{*\pm}(1430)\phi)&=&25.6\times 10^{-6}.
\end{eqnarray}
{}From the above equations, we can see that the branching ratios in S2
are about 8 times larger than those in S1. There are three main
reasons: (i) the larger decay constant in S2; (ii) contributions in
emission diagrams from the two terms $B_1$ and $B_3$ are
constructive in S2 but destructive in S1; (iii) the annihilation
diagrams could cancel the contribution from the emission diagram.
This kind of contribution in annihilation diagram is proportional to
$B_3$. The larger value for $B_3$ in S1 will results in more sizable
cancelation and the branching fractions are correspondingly reduced.

To be more explicit, we present values of the factorizable and
non-factorizable amplitudes from the emission and annihilation
topologies in Table. 1. As expected, the factorizable amplitudes
are the largest, however the annihilation magnitudes are only few times smaller
than that of factorizable emission diagrams. The non-factorizable
amplitudes  are down by a power of $\bar\Lambda/M_B\sim 0.1$
compared to the factorizable ones. The cancelation between the
twist-2 and twist-3 contributions makes them even smaller. We
demonstrate the importance of penguin enhancement in the Table. 1. It
has been known that the RG evolution of the Wilson coefficients
$C_{4,6}(t)$ dramatically increases as $t<m_b/2$, while that of
$C_{1,2}(t)$ almost remains
constant \cite{Buras}. %The phenomena has been detected in the decay
%$B\to K\pi$ \cite{Lu:2000em} and $B\to \phi
%K(K^{*})$\cite{Chen:2001pr,Li:2004ti} which play important roles in
%explaining "$K\pi$ Puzzle" and "polarization anomaly" in PQCD
%approach.

\begin{table}\caption{ Decay amplitudes for $B\to K_0^{*+}(1430)\phi$ ($\times 10^{-2}~ \mbox
{GeV}^3$)}
\begin{center}
\begin{tabular}{|c c|c|c|c|c|c|c|}
\hline \hline
$B^+\to K_0^{*+}(1430)\phi$& & $F_e$ & $M_e$ & $F_a^T$ & $F_a$ &$M_a^T$ & $M_a$\\
\hline
$S1$  &   &$-13.4$&$-0.3+i 0.0$&$-1.0-i 4.0$&$8.1+i 4.0$ &$-2.8+i3.0$ &$0.2+i 0.0$ \\
\hline
$S2$  &   &$20.4$&$-0.8+i 0.9$&$0.4+i 0.8$&$-7.1-i 12.0$ &$9.3+i2.1$ &$-0.3-i 0.2$  \\
\hline \hline $B^0\to K_0^{*0}(1430)\phi$& & $F_e$ & $M_e$ & $F_a^T$ & $F_a$&$M_a^T$ & $M_a$\\
 \hline
$S1$   &   &$-13.4$&$ -0.3+i 0.0$&$0$&$8.3+i4.0$ & $0$&$0.2-i0.1$ \\
\hline
$S2$   &    &$20.4$&$ -0.8+i 0.9$&$0$&$-7.2-i12.2$ & $0$&$-0.5-i0.3$ \\
\hline \hline
\end{tabular}\label{amp}
\end{center}
\end{table}

%
%The discussion and comparison with $B\to K^*_0\pi$ decays.

In both scenarios, the branching ratio of $B^+\to
K_0^{*+}(1430)\phi$ is a bit larger than that of $B^0\to
K_0^{*0}(1430)\phi$, and the difference is from the tree contribution in $B^+\to
K_0^{*+}(1430)\phi$. Since there exists interference between tree
and penguin diagrams in the charged channel, the direct $CP$ asymmetry appears.
So, we get the $CP$ asymmetry of $B^\pm\to K_0^{*\pm}(1430)\phi$ in
the different scenarios as follows:
\begin{eqnarray}
{\cal A}_{dir}(B^\pm\to K_0^{*\pm}(1430)\phi)&=&1.6\%,~~~~~~~~~~~~~~~S1 \nonumber\\
{\cal A}_{dir}(B^\pm\to
K_0^{*\pm}(1430)\phi)&=&1.9\%.~~~~~~~~~~~~~~~S2
\end{eqnarray}
As the neutral channel as concerned, there is no $CP$ asymmetry as
only penguin operators contribute to this channel.

%%%%%%%%%%%%%%%%%%%%%%%%%%%%%%%%%%%%%%%%%%%%%%%%
\begin{figure}[tb]
\begin{center}
\psfig{file=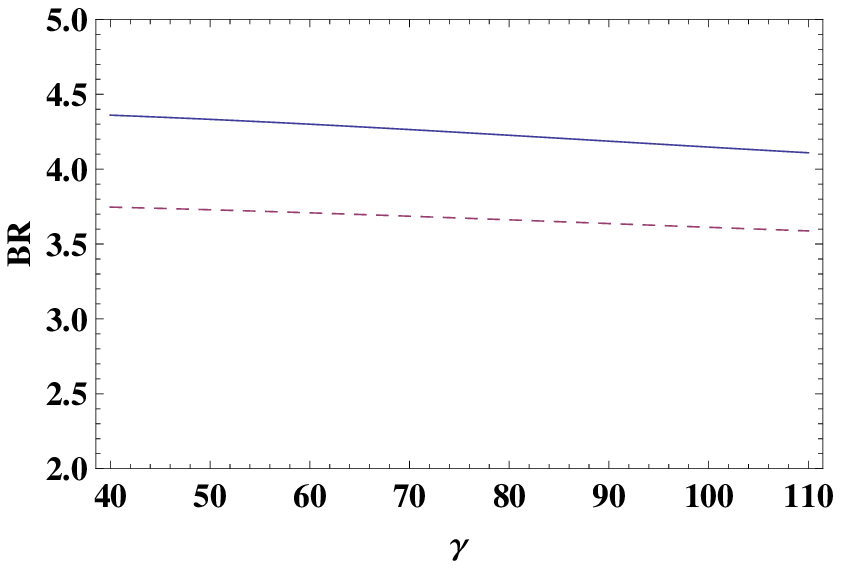,width=8.0cm,angle=0}
\psfig{file=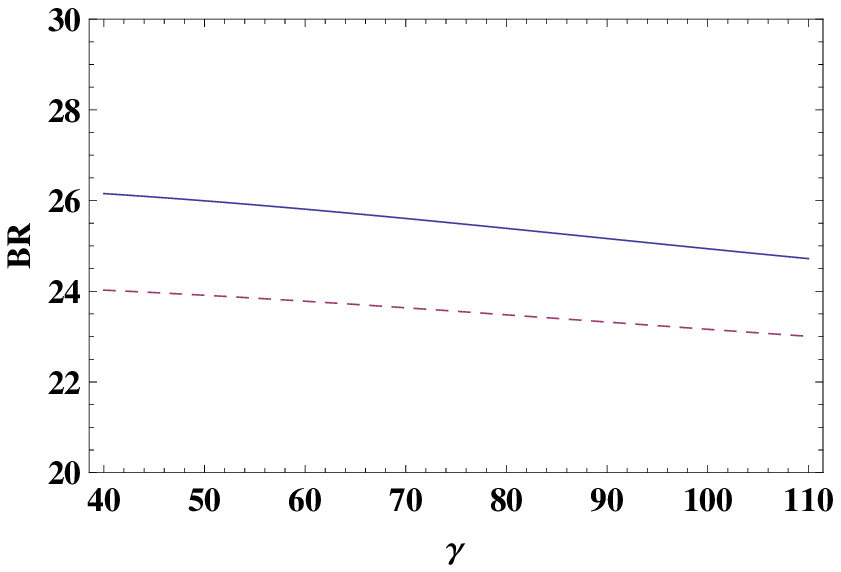,width=8.0cm,angle=0}\\
\end{center}
\caption{The dependence of the branching ratios($\times 10^{-6}$) for $B \to
K_0^*(1430)\phi$ on the CKM angle $\gamma$, where the solid (dashed)
curve is for charged (neutral) channel. The left (right) panel is
plotted in S1(S2) scenario.}\label{Feyn:O1}
\end{figure}
%%%%%%%%%%%%%%%%%%%%%%%%%%%%%%%%%%%%%%%%%%%%%%%%%%%%%
%%%%%%%%%%%%%%%%%%%%%%%%%%%%%%%%%%%%%%%%%%%%%%%%
\begin{figure}[tb]
\begin{center}
\psfig{file=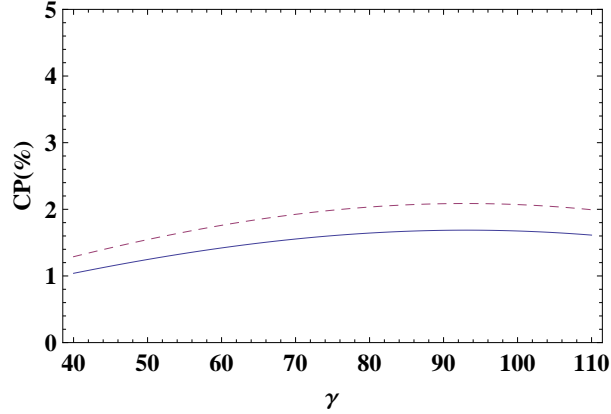,width=8.0cm,angle=0}
\end{center}
\caption{The dependence of the $CP$ asymmetry for $B^\pm \to
K_0^*{\pm}(1430)\phi$ on the CKM angle $\gamma$, where the solid
(dashed) curve is for S1 (S2) scenario}\label{Feyn:O2}
\end{figure}
%%%%%%%%%%%%%%%%%%%%%%%%%%%%%%%%%%%%%%%%%%%%%%%%%%%%%

Although we set $\gamma=70^\circ$ in the above discussions, it is
not measured accurately. In the following, we choose $\gamma$ as a
free parameter and plot the branching ratios as a function of the
angle $\gamma$ in both S1 and S2, as shown in the Fig. 2 and Fig. 3.
As seen from the figures, we note that both the branching ratios and
the $CP$ asymmetries in different scenarios are not sensitive to the
phase $\gamma$. In the decay mode $B^\pm\to K_0^{*\pm}(1430)\phi$,
the tree contribution only appears in the annihilation diagrams,
which are suppressed compared with the emission diagrams. Moreover,
the CKM element $|V_{ub}V_{us}|$ of tree diagrams is smaller than
$|V_{tb}V_{ts}|$ of penguin diagrams. {}From this point of view, we
can understand why the branching ratios and the $CP$ asymmetries are
not sensitive to the $\gamma$.

In our calculation, the major uncertainties come from our lack of
information about the scalar meson and heavy meson, involving the
decay constants and the distribution amplitudes. The latter can be
fitted from the well measured channels such as $B\to \pi\pi,K\pi$,
the scalar one is not well ascertained. These uncertainties from the
scalar meson can give sizable effects on the branching ratio, but
the $CP$ asymmetries are less sensitive to these parameters. In this
work, for instance, the twist-3 distribution amplitudes of the
scalar mesons are taken as the asymptotic form, which may give large
uncertainties. The characters of the scalar mesons need to be
studied in future work. The another uncertainty comes from the
sub-leading order contributions in PQCD approach, which have also
been neglected in the calculation. In Ref. \cite{Li:2005kt}, parts
of sub-leading order of $B\to \pi\pi,\pi K$ have been calculated,
and the results show that corrections can change the penguin
dominated processes, for example, the quark loops and
magnetic-penguin correction decrease the branching ratio of $B\to
\pi K$ by about $20\%$. We expect the similar size of uncertainty in
the decays we  analyzed , since they are also dominated by the
penguin operators.

Here we give the results with the uncertainties as follows:
 \begin{eqnarray}
  {\cal B}(B^0\to K_0^{*0}(1430)\phi)&=&(3.7^{+0.8+0.1+3.7}_{-0.7-0.1-1.7})\times 10^{-6},\nonumber\\
{\cal B}( B^-\to
K_0^{*-}(1430)\phi)&=&(4.3^{+0.9+0.1+4.3}_{-0.8-0.1-2.0})\times
10^{-6}
\,\,\,\,\,\,\,\,\,\,\,\,S1;\nonumber\\
{\cal B}(B^0\to K_0^{*0}(1430)\phi)&=&(23.6^{+5.6+0.8+10.9}_{-5.0-0.6-5.8})\times 10^{-6},\nonumber\\
{\cal B}( B^-\to
K_0^{*-}(1430)\phi)&=&(25.6^{+6.2+0.9+12.1}_{-5.4-0.8-6.5})\times
10^{-6}  \,\,\,\,\,\,\,\,\,\,S2.
\end{eqnarray}
In the above results, the first uncertainty comes from the decay
constants, and the second one is from the uncertainties of B1 (B3) in
the amplitude distributions of the scalar meson. The last one comes from the uncertainty in
the $B$ meson shape parameter $\omega=(0.40\pm0.05)$ GeV. This kind
of uncertainties is extremely large. The change of the shape
parameter will mainly affect the emission diagram including the
$B\to K^{*}_0$ form factor while the annihilation diagram,
especially factorizable diagram, will not be affected sizably.
Remember that the annihilation diagram could cancel part of
contributions from emission diagram and thus the branching fractions
are sizably changed due to the shape parameter.

In the QCD factorization approach, the results are listed as
\cite{Cheng:2007st}:
 \begin{eqnarray}\label{results-qcdf}
{\cal B}(B^0\to K_0^{*0}(1430)\phi)&=&(0.9^{+0.3+0.4+19.3}_{-0.3-0.3-0.5})\times 10^{-6},\nonumber\\
{\cal B}(B^-\to
K_0^{*-}(1430)\phi)&=&(1.0^{+0.3+0.4+20.2}_{-0.3-0.3-0.5})\times
10^{-6}
\,\,\,\,\,\,\,\,\,\,\,\,S1;\nonumber\\
{\cal B}(B^0\to K_0^{*0}(1430)\phi)&=&(16.9^{+6.2+1.7+51.8}_{-4.7-1.6-12.0})\times 10^{-6},\nonumber\\
{\cal B}( B^-\to
K_0^{*-}(1430)\phi)&=&(17.3^{+6.2+1.7+52.4}_{-4.7-1.7-12.1})\times
10^{-6},\,\,\,\,\,\,\,\,\,S2.
\end{eqnarray}
Comparing two group of results, we note that our central values are
much large than the results from QCDF in both two scenarios. It is
mostly because that the form factor derived from Eq.~(\ref{fe}) is
larger than $F_1^{B\to \ks}(q^2=0)=0.21~(0.26)$ used in QCDF, which is
calculated under S1 (S2) scenario in the covariant light-front model \cite{Cheng:2003sm}.
In addition, our results suffer from contribution from the
annihilation diagrams, as demonstrated in the Table. 1. In fact, the
contribution from annihilation can take the major uncertainties in
the QCDF, as shown in the Eq. (\ref{results-qcdf}).

In the S1, for the neutral channel, our result is agree with
experimental data well, but the result of the charged one is smaller
than the data, though it is consistent within theoretical
uncertainties. In the S2, both results are much larger than the
data. The predictions in both scenarios suffer from very large
uncertainties from the hadronic input parameters. Fortunately, most
of these uncertainties will cancel out when we consider the ratio of
branching fractions. It is convenient to define the ratio
\begin{eqnarray}
 R&=& \frac{\tau (B^0)}{\tau (B^+)}\frac{{\cal B}(B^{\pm}\to \phi K^{*\pm}_0)}{{\cal B}(B^0\to \phi
 K^{*0}_0)},
\end{eqnarray}
which is predicted as
\begin{eqnarray}
 R&=&1.08 \pm 0.01 ,\,\,\,\,\,\,\,\,\,\,\,\,S1;\nonumber\\
 R&=& 1.01\pm0.01.\,\,\,\,\,\,\,\,\,\,\,\,S2;
\end{eqnarray}
Using the two experimental results, one can easily obtain the
experimental data for this ratio
\begin{eqnarray}
 R_{\rm exp}=1.68\pm0.51,
\end{eqnarray}
where all uncertainties are added in quadrature. For this ratio, the
uncertainties from theoretical predictions are small while the
experimental data has large uncertainties.

As a summary, we have studied the hadronic charmless decay mode $B \to
K_0^*(1430)\phi$ within the framework of perturbative QCD approach
in the standard model. Under two different scenarios, we explored
the branching ratios and related $CP$ asymmetries. We find that
besides the dominant contributions from the factorization emission
diagrams, the penguin operators in annihilation can change the ratio
remarkably. The central value of our results are larger than those from QCD
factorization. Compared with experimental data from BaBar, in the
S1, the result of neutral channel is agree with experimental data
well, but the result of the charged one is a bit smaller than the
data, though it is consistent within theoretical uncertainties. In
the S2, both results are much larger than the data but the
uncertainties are typically large. The ratio of branching fractions
is found to have small uncertainties in the theoretical side.

\section*{Acknowledgments}
The work of C.S.K. was supported in part by Basic Science Research
Program through the NRF of Korea funded by MOEST (2009-0088395) and
in part by KOSEF through the Joint Research Program
(F01-2009-000-10031-0). The work of Ying Li was supported by the
Brain Korea 21 Project and by the National Science Foundation under
contract Nos.10805037 and 10735080.

%-------------------------------------------------------------------------------
%                               references
%--------------------------------------------------------------------------------

\end{document}